\documentstyle[12pt]{article}
\newlength{\bredde}
\def\slash#1{\settowidth{\bredde}{$#1$}\ifmmode\,\raisebox{.15ex}{/}
\hspace*{-\bredde} #1\else$\,\raisebox{.15ex}{/}\hspace*{-\bredde} #1$\fi}
\newcommand{\beq}{\begin{equation}}
\newcommand{\eeq}{\end{equation}}
\newcommand{\ba}{\begin{array}{ccc}}
\newcommand{\ea}{\end{array}}

\newcommand{\noi}{\vspace{12pt}\noindent}
\newcommand{\lG}{\raise.3ex\hbox{$\stackrel{\leftarrow}{G}$}}
\newcommand{\lU}{\raise.3ex\hbox{$\stackrel{\leftarrow}{U}$}}
\newcommand{\lP}{\raise.3ex\hbox{$\stackrel{\leftarrow}{{\cal P}}$}}
\newcommand{\leta}{\raise.3ex\hbox{$\stackrel{\leftarrow}{\eta}$}}
\newcommand{\lOmega}{\raise.3ex\hbox{$\stackrel{\leftarrow}{\Omega}$}}
\newcommand{\ldr}{\raise.3ex\hbox{$\stackrel{\leftarrow}{\delta^r}$}}

\def\m2{{\mathcal{M}}^{\dagger}{\mathcal{M}}}
\def\mb2{M^2}

\def\beqn{\begin{eqnarray}}
\def\eeqn{\end{eqnarray}}

\def\gtwid{\raise.3ex\hbox{$>$\kern-.75em\lower1ex\hbox{$\sim$}}}
\def\ltwid{\raise.3ex\hbox{$<$\kern-.75em\lower1ex\hbox{$\sim$}}}

\begin{document}
\topmargin -1.4cm
\oddsidemargin -0.8cm
\evensidemargin -0.8cm
\title{\Large{{\bf Some remarks on the ``classical'' large $N$ limit}}}

\vspace{1.5cm}

\author{~\\{\sc Poul Olesen} \\~\\
The Niels Bohr Institute\\ Blegdamsvej 17\\ DK-2100 Copenhagen {\O}\\
Denmark}
\date{\today} 
\maketitle
\vfill
\begin{abstract} 
It has been proposed some time ago that the large $N-$limit can be understood
as a ``classical limit'', where commutators in some sense approach the
corresponding Poisson brackets. We discuss this in the light of 
some recent numerical
results for an SU($N$) gauge model, which do not agree with this
``classicality'' of the large $N-$limit. The world sheet becomes very
crumpled. We speculate that this effect would disappear in supersymmetric
models. 
\end{abstract}
\vfill

\begin{flushleft}
{\it To Holger Bech Nielsen on his 60th birthday}
\end{flushleft}
\thispagestyle{empty}
\newpage

\noi

In the 1970ies Holger and I discussed the confinement problem quite a lot.
In those days the picture of confinement by means of a string connecting
the quarks became more and more accepted.
It was known that the Nambu-Goto action could be written as a square root of
the Poisson bracket squared. One of the subjects we discussed was 
whether somehow the Poison bracket was related to the commutator like
in quantum mechanics, in some limit. However,
we did not gain any insight in this. After the work by Eguchi and Kawai
it became of course clear that the relevant limit was the large $N$
limit, where the action could be expressed in terms of commutators. 
Later on Hoppe \cite{hoppe} constructed a very suggestive relation between
Poisson brackets and commutators. This was then used in ingenious ways
by a number of authors \cite{floratos,fairlie,bars}, in connection
with the Eguchi-Kawai formulation of the large $N$ theory.

\noi
 
To give a brief review of this approach,
let us consider SU($N$) with the gauge field $A_\mu$ and the generators
$l_{\bf k},{\bf k}=(k_1,k_2)$,
\beq
(A_\mu)^j_i=\sum_{\bf k}a^{\bf k}_\mu~ (l_{\bf k})^j_i,
\eeq
where $a^{\bf k}_\mu$ are expansion coefficients to be integrated in functional
integrals. This expansion can be compared to the one for a string variable 
$X_\mu (\sigma,\tau)$,
\beq
X_\mu (\sigma,\tau)=\sum_{\bf k}a^{\bf k}_\mu~e^{i{\bf k\sigma}},
\eeq
with $\sigma=(\sigma,\tau)$. The variables $X_\mu$ are the Weyl
transform of the matrices $A_\mu$.
The generators can be constructed explicitly
in terms of the Weyl matrices, and have the commutation relation
\beq
[l_{{\bf k}_1},l_{{\bf k}_2}]=i\frac{N}{2\pi}~\sin \left(\frac{2\pi}{N} 
{\bf k}_1\times {\bf k_2}\right)~l_{{\bf k}_1+{\bf k}_2}\rightarrow i
\left({\bf k}_1\times {\bf k_2}\right)~l_{{\bf k}_1+{\bf k}_2}
\label{sine}
\eeq
for $N\rightarrow \infty$, where ${\bf k}_1\times {\bf k_2}=
(k_1)_1(k_2)_2-(k_2)_1(k_1)_2$. This leads to a comparison
of the commutator with the Poisson bracket. For example, one has
\beq
[A_\mu,A_\nu]=i\frac{N}{2\pi}\sum_{\bf k,p}a^{\bf k}_\mu~ a^{\bf p}_\nu~
\sin \left(\frac{2\pi}{N}{\bf k}\times {\bf p}\right)~
l_{{\bf k}_1+{\bf k}_2},
\label{modes}
\eeq
and
\beq
\{X_\mu (\sigma),X_\nu (\sigma)\}=i\sum_{\bf k,p}a^{\bf k}_\mu~ a^{\bf p}_\nu~
({\bf k}\times{\bf p})~e^{i({\bf k}+{\bf p})}.
\eeq
Using 
\beq
{\rm tr}~ l_{\bf k}l_{\bf p}\propto \delta_{\bf k+p,0},~~\int d^2\sigma~
e^{i({\bf k}+{\bf p}){\bf \sigma}}\propto \delta_{\bf k+p,0},
\eeq
one can derive that
\beq
{\rm tr}~[A_\mu,A_\nu]^2\rightarrow {\rm const.}~\int d^2\sigma 
\{X_\mu,X_\nu\}^2,
\eeq
where the constant (depending on $N$) can be removed by a suitable
normalization, and where it was used that
\beq
\frac{N}{2\pi}~\sin \left(\frac{2\pi}{N} {\bf k}_1\times {\bf k_2}\right)~
\rightarrow {\bf k}_1\times {\bf k_2}
\eeq
for $N\rightarrow\infty$, even {\it inside} the relevant sum over modes. In 
this sense the commutator approaches the  Poisson 
bracket \cite{floratos,fairlie,bars}. 

\noi

Unfortunately this argument, leading to a nice connection between strings and
fields in the large $N-$limit, depends on performing the limit $N\rightarrow
\infty$ inside sums over modes like in eq. (\ref{modes}). This is
evidently valid if the low modes dominate, since the difference
between sin $x$ and the linear function $x$ is considerable when $x$ is not 
close to 0.

\noi 

Recently Anagnostopoulos, Nishimura and the author \cite{us} started to
investigate numerically the mode-distribution for large $N$ in a
four-dimensional SU($N$) model with the partition function
\beq
\int dA_\mu~\exp\left(\frac{1}{4g^2}{\rm tr}[A_\mu,A_\nu]^2\right).
\label{action}
\eeq
This model has been shown to exist \cite{krauth,hotta,austing}.
The expectation value of the commutator squared 
\beq
M=<{\rm tr}[A_\mu,A_\nu]^2>,
\label{1}
\eeq
can be found from a scaling argument \cite{hotta} to be $N^2-1$, and we 
checked our
numerical method by seeing that it leads very precisely to this
result. 

Then we computed the corresponding expectation value of the Poisson
bracket by introducing the Weyl-transform\footnote{The 
Weyl transform $X_\mu(\sigma)$ of the matrix $A_\mu$ is given by $X_\mu(\sigma)
\propto\sum_k \exp (ik\sigma)~$tr$(l_kA_\mu)$}$X_\mu (\sigma)$ (defined
on a torus) of $A_\mu$, and we therefore found the behavior of
\beq
<P>=<\int d^2\sigma~\{X_\mu(\sigma),X_\nu (\sigma)\}^2>.
\eeq
Here the normalization is such that if only the very lowest modes are kept,
one has the result $<M>=<P>$. For the details of this construction, we
refer to the paper \cite{us}. One important point discussed 
in this paper is that the Poisson bracket is gauge dependent, and hence
$<P>$ is in general gauge dependent, in contrast to the gauge invariant
$<M>$. We have selected a gauge fixing analogous to the Landau gauge.
Of course, one can anyhow argue that if the commutator really approaches
the Poisson bracket, gauge invariance of the former implies gauge
invariance of the latter, at least approximatively.

\noi

In our numerical calculations we took $N=15,25,$ and 35. The results were 
quite embarrasing. Keeping only a few modes, we got $<M>\approx <P>$, but 
including all modes these two quantities differ considerably: Whereas very 
precisely we find $<M>=N^2-1$,
it turns out that $<P>$ rather grows like $O(N^4)$. Therefore the two
expectation values do not at all agree, and the discrepancy actually
increases with $N$. We can therefore say that in the model with the
action (\ref{action}), the Poisson bracket is not approximated by
the commutator. 

We also computed the average of the area of the world sheet,
\beq
<A>=<\int d^2\sigma \sqrt{\{X_\mu (\sigma),X_\nu (\sigma)\}^2}~>.
\eeq
It turns out that 
$<A>\approx <M>$. Thus, the expectation value of the Nambu-Goto action
does indeed behave approximately as the commutator. This does, however, not
show that the action (\ref{action}) is approximately equal to the Nambu-Goto
action, since we have computed only one moment of $A$.

\noi

The negative result given above shows that the question of
whether the large $N$ limit is a classical limit, depends on the dynamics, and
a simple model like (\ref{action}) is not able to suppress the higher
modes, and hence one cannot use the limit (\ref{sine}) inside sums over
modes. The string picture resulting from the behavior of $X_\mu (\sigma)$
turns out to be an extremely crumpled string, with an almost infinite
($\approx 33$) Hausdorff dimension. This is another way of seeing that 
higher modes are important, since dominance of lower modes would lead to 
a smooth string. 

\noi

As pointed out in \cite{us} the model (\ref{action}) does exist in a
world-sheet version, since the commutator can be replaced by star products
of the $X'$s, so the action can be rewritten as
\beq
\int dX_\mu~\exp\left[\frac{-1}{4g^2N}\int d^2\sigma \left(X_\mu (\sigma)
\star X_\nu (\sigma)-X_\nu (\sigma)\star X_\mu (\sigma)\right)^2\right],
\eeq
which exists because the corresponding matrix model exists. The world sheet
defined by this theory (with an infinite number of derivatives in the
star product) is the same as before, and is thus extremely crumpled.

\noi

Thus the dynamical question of the existence of a ``classical'' large 
$N-$limit boils down to finding a rather smooth string, since any
``violent'' string would require higher modes. Presumably the QCD
string is smooth. In this case the action (\ref{action}) is replaced by
a quenched action, which hopefully can manage to suppress the high
modes. Another possibility is to take a supersymmetric version of
(\ref{action}), since supersymmetry in general makes strings more smooth:
For example, in the vacuum energy the tachyon is removed by supersymmetry.
For the simple bosonic string it was shown long time ago by Alvarez
\cite{alvarez} that the bosonic string is very crumpled,
and by the author \cite{po} that the tachyon produces this crumpling. 
Alvarez' formula for the free energy \cite{alvarez},
\beq
F=\frac{1}{2\pi \alpha '}T\sqrt{R^2-R_c^2},
\eeq
can actually be interpreted \cite{po} as a formula for the lowest tachyonic 
energy, with the tachyon mass squared being proportional to $-R_c^2$.
The phenomenon of crumpling should thus not occur in the supersymmetric case. 
We hope that these questions will be investigated numerically for
supersymmetric versions of (\ref{action}) in the future.

\noi
Finally, I would like to take this opportunity to thank Holger for
pleasant years of discussions of very many aspects of physics.

\end{document}